\RequirePackage{lineno_babar_revtex}
\documentclass[twocolumn,showpacs,aps,preprintnumbers,amsmath,amssymb,floatfix,linenumbers]{revtex4}

\usepackage{lineno_babar_revtex}

\usepackage{graphicx}
\usepackage{color}
\usepackage{dcolumn}     

\newcommand{\BaBarPubYear}       {12}

\newcommand{\BaBarPubNumber}     {027}
\newcommand{\SLACPubNumber}   {16291}

\input babarsym
\newcommand{\bi}{\begin{itemize}}
\newcommand{\ei}{\end{itemize}}
\newcommand{\ben}{\begin{enumerate}}
\newcommand{\een}{\end{enumerate}}
\newcommand{\bc}{\begin{center}}
\newcommand{\ec}{\end{center}}
\newcommand{\bt}{\begin{table}}
\newcommand{\et}{\end{table}}
\newcommand{\be}{\begin{equation}}
\newcommand{\eeq}{\end{equation}}
\newcommand{\ba}{\begin{eqnarray}}
\newcommand{\ea}{\end{eqnarray}}

\newcommand{\la}{\ifmmode {\leftarrow} \else {$\leftarrow$}\fi}
\newcommand{\Ra}{\ifmmode {\Rightarrow} \else {$\Rightarrow$}\fi}
\newcommand{\La}{\ifmmode {\Leftarrow} \else {$\Leftarrow$}\fi}
\newcommand{\Lra}{\ifmmode {\Longrightarrow} \else {$\Longrightarrow$}\fi}
\newcommand{\Lla}{\ifmmode {\Longleftarrow} \else {$\Longleftarrow$\fi}}
\newcommand{\Llra}{\ifmmode {\Longleftrightarrow} \else {$\Longleftrightarrow$\fi}}
\newcommand{\Lk}{\ifmmode {{\cal L}} \else {${\cal L}$}\fi}
\newcommand{\Wt}{\ifmmode {{\cal W}} \else {${\cal W}$}\fi}
\newcommand{\Br}{\ifmmode {{\cal B}} \else {${\cal B}$}\fi}
\newcommand{\N}{\ifmmode {{\cal N}} \else {${\cal N}$}\fi}
\newcommand{\G}{\ifmmode {{\cal G}} \else {${\cal G}$}\fi}
\newcommand{\E}{\ifmmode {{\cal E}} \else {${\cal E}$}\fi}
\newcommand{\Pfr}{\ifmmode {{\cal F}} \else {${\cal F}$}\fi}
\newcommand{\Aone}{\ifmmode {{\cal A}_1} \else {${\cal A}_1$}\fi}
\newcommand{\rha}{\ifmmode{\mbox{\rho^2_{{\cal A}_1}}} \else {\mbox{$\rho^2_{{\cal A}_1}$}}\fi}
\newcommand{\rhf}{\ifmmode{\rho^2_{\cal F}}\else{\mbox{$\rho^2_{\cal F}$}}\fi}
\newcommand{\om}{\ifmmode {w} \else {$w$}\fi}
\newcommand{\dom}{\ifmmode {\Delta w} \else {$\Delta w$}\fi}

\newcommand{\tBz}{\ifmmode {\tau_{\Bz}} \else {$\tau_{\Bz}$}\fi}
\newcommand{\tBp}{\ifmmode {\tau_{\Bu}} \else {$\tau_{\Bu}$}\fi}

\newcommand{\psoft}{\ifmmode {\pi_{s}} \else {$\pi_{s}$}\fi}
\newcommand{\plab}{\ifmmode{p} \else {$p$} \fi}
\newcommand{\ctdl}{\ifmmode{ \cos(\theta_{\Dstar\ell}) } \else {$\cos(\theta_{\Dstar\ell})$} \fi}
\newcommand{\ks}{\ifmmode{k^*} \else {$k^*$} \fi}
\newcommand{\mnutag}{\ifmmode{m^2_{\nu ,tag}} \else {$m^2_{\nu ,tag}$}\fi} 
\newcommand{\mnusig}{\ifmmode{m^2_{\nu ,sig}} \else {$m^2_{\nu ,sig}$}\fi} 
\newcommand{\DTau}{\ifmmode {\Delta \tau} \else {$\Delta \tau$}\fi}
\newcommand{\ggcc}{\ifmmode {GeV^2/c^4} \else {$GeV^2/c^4$}\fi}
\def\BpBm {\ensuremath{B^+ {\kern -0.16em \Bub}}}
\def\dT{\ensuremath{\Delta t}'}
\def\DT{\ensuremath{\Delta t}}
\def\dZ{\ensuremath{\Delta Z}}

\newcommand{\BtoDs}{\mbox{$\Bzb\rightarrow D^{*+} \ell^- \bar{\nu_\ell}$}\xspace}

\newcommand{\magqp}{\mbox{$|q/p|$}}
\def\poverq2{\ensuremath{\bigg\vert\frac{p}{q}\bigg\vert^2}\xspace}
\def\qoverp2{\ensuremath{\bigg\vert\frac{q}{p}\bigg\vert^2}\xspace}

\def\BzBz     {\ensuremath{\mbox{\Bz {\kern -0.1em \Bz}}}\xspace}
\def\BzBzb     {\ensuremath{\mbox{\Bz {\kern -0.1em \Bzb}}}\xspace}
\def\BzbBzb   {\ensuremath{\mbox{\Bzb {\kern -0.1em \Bzb}}}\xspace}
\def\mm {{\ensuremath{{{\cal M}_{\mathrm{miss}}}^2}}\xspace}
\newcommand{\Brec}{\mbox{$B_{R}$}}
\newcommand{\Btag}{\mbox{$B_{T}$}}

\def\Kp{\ensuremath {K^{+}\xspace}}
\def\Km{\ensuremath {K^{-}\xspace}}
\def\Kt{\ensuremath {K_{tag}\xspace}}
\def\Kr{\ensuremath {K_{rec}\xspace}}

\def\At{\ensuremath{\mathcal{A}_{K}}\xspace}
\def\Ar{\ensuremath{\mathcal{A}_{r\ell}}\xspace}
\def\Are{\ensuremath{\mathcal{A}_{rec,e}}\xspace}
\def\Arm{\ensuremath{\mathcal{A}_{rec,\mu}}\xspace}
\def\All{\ensuremath{\mathcal{A}_{CP}}\xspace}

\def\Kt{\ensuremath{K_{T}}\xspace}
\def\Kr{\ensuremath{K_{R}}\xspace}

\def\all{\ensuremath{\mathcal{A}_{CP}}}
\def\tbz{\ensuremath{\tau_{\Bz}}}
\def\dCP{\ensuremath{\Delta_{CP}}}

\def\dZ{\ensuremath{\Delta Z}}
\def\Zr{\ensuremath{Z_{\mathrm{rec}}}}
\def\Zt{\ensuremath{Z_{\mathrm{tag}}}}
\def\qoverp{\ensuremath{\frac{q}{p}}}

\long\def\inst#1{\par\nobreak\kern 4pt\nobreak
    {\it #1}\par\vskip 10pt plus 3pt minus 3pt}

\def\babar{\mbox{\slshape B\kern-0.1em{\smaller A}\kern-0.1em
    B\kern-0.1em{\smaller A\kern-0.2em R}}}

\begin{document}

\title{{Search for {\boldmath\CP} violation in {\boldmath \Bz-\Bzb} mixing
using partial reconstruction of
{\boldmath{$\Bzb \rightarrow D^{*+} X \ell^{-} \bar{\nu}_{\ell}$}}
and kaon tagging}}
%
\author{J.~P.~Lees}
\author{V.~Poireau}
\author{V.~Tisserand}
\affiliation{Laboratoire d'Annecy-le-Vieux de Physique des Particules (LAPP), Universit\'e de Savoie, CNRS/IN2P3,  F-74941 Annecy-Le-Vieux, France}
\author{E.~Grauges}
\affiliation{Universitat de Barcelona, Facultat de Fisica, Departament ECM, E-08028 Barcelona, Spain }
\author{A.~Palano$^{ab}$ }
\affiliation{INFN Sezione di Bari$^{a}$; Dipartimento di Fisica, Universit\`a di Bari$^{b}$, I-70126 Bari, Italy }
\author{G.~Eigen}
\author{B.~Stugu}
\affiliation{University of Bergen, Institute of Physics, N-5007 Bergen, Norway }
\author{D.~N.~Brown}
\author{L.~T.~Kerth}
\author{Yu.~G.~Kolomensky}
\author{M.~J.~Lee}
\author{G.~Lynch}
\affiliation{Lawrence Berkeley National Laboratory and University of California, Berkeley, California 94720, USA }
% ...
\collaboration{The \babar\ Collaboration}
\noaffiliation

\begin{flushleft}
{
\babar-PUB-\BaBarPubYear/\BaBarPubNumber \\
SLAC-PUB-\SLACPubNumber    \\  
}
\end{flushleft}

\begin{abstract}
We describe in detail a {previously} published measurement of \CP violation {in} \Bz-\Bzb\
oscillations, based on 
an integrated luminosity of 425.7\invfb
collected by the \babar\
experiment at the PEPII collider.
We apply a novel technique to a sample of about 6 million \BtoDs\ 
decays 
selected with partial reconstruction of the \dsp\ meson. The
charged lepton identifies the flavor of one {\em B} meson at its decay time,
the flavor of the other {\em B} is determined by kaon tagging.

We determine a \CP violating asymmetry 
$\All = (N(\Bz\Bz) - N(\Bzb\Bzb))/(N(\Bz\Bz)+ N(\Bzb\Bzb))
=(0.06\pm 0.17^{+0.38}_{-0.32})\%$ 
corresponding to
$\dCP =1 - |q/p| =(0.29\pm0.84^{+1.88 }_{-1.61})\times 10^{-3}$.
This measurement is consistent and competitive with those obtained
at the {\em B} factories with dilepton events.
\end{abstract}



\maketitle

\setcounter{footnote}{0}


\section{Introduction}
\begin{linenomath}
\label{sec:intro}
The time evolution of neutral {\em B} mesons is governed by the
Schr\"odinger equation :
\begin{equation}\label{eq:sch}
-i  \frac{\partial}{\partial t} 
 \Psi
= {\cal H} 
\Psi
\end{equation}
\end{linenomath}
where $\Psi=\psi_1 |\Bz \rangle$ + $\psi_2 | \Bzb \rangle$  and $ \Bz  = (\bar{b} d) $ and $ \Bzb= (b \bar{d}) $ are
flavor eigenstates. 
The Hamiltonian  ${\cal H} =M - \frac{i}{2} \Gamma$ is the combination of two $2\times 2$ hermitian matrices,
$ M^\dagger =M ,~\Gamma^\dagger = \Gamma $, expressing dispersive
and absorptive contributions respectively.
The two eigenstates of ${\cal H}$, with well-defined values of
mass ($m_L$, $m_H$) and decay width ($\Gamma_L$, $\Gamma_H$), are expressed 
in terms of \Bz\ and \Bzb, as

\begin{linenomath}
\begin{eqnarray}
|B_L \rangle & = & p |\Bz \rangle + q |\Bzb \rangle    \\ \nonumber
|B_H \rangle & = & p |\Bz \rangle - q |\Bzb \rangle,
\end{eqnarray}
\end{linenomath}
where 
\begin{linenomath}
\begin{equation}
 \frac{q}{p} = \sqrt{\frac{M_{12}^* -i\Gamma_{12}^* / 2}
                                    {M_{12}-i\Gamma_{12}/ 2}}.
\end{equation}
\end{linenomath}
The process of \Bz-\Bzb\ flavor mixing is therefore governed by
two real parameters, $|M_{12}|,
|\Gamma_{12}|$, and by the phase $\phi_{12} = \mathrm{arg}( -
\Gamma_{12}/M_{12})$.

The value of  $|M_{12}|$ is related to the frequency of {\Bz - \Bzb\ 
oscillations}, $\Delta m$,  by the relation:
\begin{linenomath}
\begin{equation}
 \Delta m = m_H - m_L = 2|M_{12}|, 
\end{equation}
\end{linenomath}
whereas the following expression
 relates the decay width difference $\Delta \Gamma$ to $|\Gamma_{12}|$ and
 $\phi_{12}$ :
\begin{linenomath}
\begin{equation}
\Delta \Gamma = \Gamma_L - \Gamma_H = 2 |\Gamma_{12}| \cos \phi_{12}.
\end{equation}
\end{linenomath}
A third observable probing mixing is the \CP\ mixing asymmetry 
\begin{linenomath}
\begin{equation}\label{eq:A_SL}
\all = \frac{\bar{\cal P}-{\cal P}}{\bar{\cal P}+{\cal P}} \simeq
2(1 -  | \frac{q}{p}|) =
\frac{\Delta \Gamma}{\Delta m} \tan\phi_{12} ,
\end{equation}
\end{linenomath}
where ${\cal P} = \mathrm{prob} (\Bz \rightarrow \Bzb)$ is the probability that
a state, produced as a \Bz, decays as a \Bzb, $\bar{\cal P} = \mathrm{prob} (\Bzb \rightarrow \Bz)$ is
the probability for the \CP\ conjugate oscillation, the second equality holds if $|q/p|\simeq 1$, and the last if $|\Gamma_{12}/M_{12}| \ll 1$.

In the Standard Model the dispersive term $M_{12}$ is dominated by box
diagrams involving two top quarks. Owing to the large top mass, a sizeable
value of $\Delta m$ is expected. The measured value {$\Delta m = 0.510 \pm 0.004$
$\mathrm{ps}^{-1}$~\cite{PDG}} is consistent with the SM expectation. The period
corresponds to about eight times the \Bz\ average lifetime.

As only the few final states common to \Bz\ and \Bzb\ contribute to 
$|\Gamma_{12}|$, small values of $\Delta \Gamma$ and \all\ are
expected. 
One of the most recent theoretical calculations
based on the SM~\cite{SM},
including  NLO QCD correction, predicts :
\begin{linenomath}
\begin{eqnarray}
\dCP = 1 -|q/p|  {\simeq}  \frac{1}{2} \all = -(2.4^{+0.5}_{-0.6}) \times 10^{-4}.
\end{eqnarray}
\end{linenomath}
Sizeable deviations from zero would therefore be a clear indication of New
Physics.
A detailed review of possible NP contributions to \CP-violation in {\Bz - \Bzb\ }
mixing can be found in \cite{Lenz}. In this paper, we describe the
measurement of \all\ performed by the \babar\ collaboration with a
novel technique, {previously} published in~\cite{babar_prl}, which, due to the 
analysis complexity, requires a more detailed description.

This article is organized as follows.
An overview of the current experimental situation and the strategy of this measurement are
reported in Sec.~\ref{sec:rev}.
The \babar\ detector is described briefly in Sec.~\ref{sec:det}. Event
selection and sample composition are then described in
Sec.~\ref{sec:evsel}. Tagging the flavor of the {\em B} meson is described in Sec.~\ref{sec:ktag}.
The measurement of \All\ is described in
Sec.~\ref{sec:fit}, the fit validation is described in
Sec.~\ref{sec:val}, 
the discussion of the systematic uncertainties
follows in Sec.~\ref{sec:syst}, while we summarize the results and draw
our conclusions in Sec.~\ref{sec:res} and \ref{sec:conc}.

\section{ Experimental overview and description of the measurement}
\label{sec:rev}
In hadron collider experiments, $b\bar{b}$ pairs produced at the parton level 
hadronize generating the {{\em b} hadrons}, which eventually
decay weakly.
In {\em B} factories, pairs of opposite flavor {\em B}-mesons are produced through the process \mbox{$\epem
  \rightarrow \Upsilon(4S) \rightarrow B \bar{B}$} in an entangled quantum
state. Because of flavor mixing, decays of two $\Bz$ or
$\Bzb$ mesons are observed.
If \CP is violated in mixing, ${\cal P} \neq \bar{\cal P}$ and
 a different number of \BzBz\ events with respect 
to \BzbBzb\ is expected. The asymmetry is measured by selecting flavor tagged final
states $f$, for which the decay $\Bz \rightarrow f$ is allowed and {the decay}
$\Bzb \rightarrow f$ is forbidden. Inclusive semileptonic decays $\Bz
\rightarrow \ell^+ \nu_\ell X$ have been used in the past, due to the
large branching fraction and high selection efficiency (unless the contrary is explicitly
stated, we always imply charge conjugated processes; ``lepton'' $\ell$
means
either electron or muon). 
Assuming $CPT$ symmetry for semileptonic decays
($\Gamma(\Bz
\rightarrow \ell^+ \nu_\ell X) = \Gamma(
\Bzb
\rightarrow \ell^- \bar{\nu_\ell} \bar{X})$), the observed asymmetry 
is directly related to \CP\ violation in mixing:
\begin{linenomath}
\begin{equation}
\frac{{\cal N}(\ell^+\ell^+)-{\cal N}(\ell^-\ell^-)}
{{\cal N}(\ell^+\ell^+)+{\cal N}(\ell^-\ell^-)}=\all
\end{equation}
\end{linenomath}
 where ${\cal N}(\ell^\pm\ell^\pm)$ is the efficiency-corrected 
number of equal charge dilepton events 
after background subtraction.

Published results from { CLEO~\cite{CLEO} and} the {\em B} factory experiments 
Belle~\cite{Belle_dilep} and \babar~\cite{BaBar_dilep, BaBar_dilep2}, 
based on the analysis of dilepton events,
are consistent with the SM expectation.
The $D\emptyset$ collaboration~\cite{D0_mumu}, using a dimuon sample, 
obtained a more precise
measurement, which however includes
contributions from both \Bz\ and {\Bs\ } mixing. 
They observe a deviation larger than three
standard deviations from the SM expectation. 
Measurements based  on the reconstruction of \mbox{ ${ \Bsb}
\rightarrow D_s^{(*)+} \ell \bar{\nu_\ell}$} decays~\cite{D0_Bs, LHCB_Bs}
and of  \mbox{ $\Bzb
\rightarrow D^{(*)+} \ell \bar{\nu_\ell}$} decays~\cite{D0_Bd, LHCB_Bd} are
compatible both with
the SM and with $D\emptyset$.

The dilepton measurements benefit from the large number of events that
can be selected at {\em B} factories or at hadron colliders. 
However, they rely on
the use of control samples to subtract the charge-asymmetric
background originating from
hadrons wrongly identified as leptons
or leptons from light hadron
decays, and to compute the charge-dependent lepton identification
asymmetry that may produce a false signal. The systematic
uncertanties associated with the corrections for these effects 
constitute a severe limitation of the precision 
of the measurements.
Particularly obnoxious is the case when a lepton from
a direct $B$ semileptonic decay is combined with a lepton of equal charge from a charm
meson produced in the decay of the other $B$. As the mixing probability
is rather low, this background process is enhanced with respect to the signal, so
that stringent kinematic selections need to be applied. Authors
of  \cite{Bori} suggest that at least a part of the $D\emptyset$
dilepton discrepancy could be due 
to charm decays.

Herein we present in detail a measurement which overcomes
these difficulties with  a new approach.
To reduce the background dilution from \Bp\Bm\ or from light quark events, we reconstruct
\BtoDs\ decays  with a very efficient selection using
only the charged lepton and the low-momentum pion (\psoft) from the 
$\dsp \rightarrow \Dz \psoft$ decay. A state decaying as a \Bz (\Bzb)
meson produces $\ell^+\psoft^-$ ($\ell^-\psoft^+$). 
We use charged kaons from decays of the other \Bz\ to tag its
flavor (\Kt). Kaons are mostly produced in the Cabibbo-favored (CF) process
$\Bz \rightarrow \bar{D}X, \bar{D} \rightarrow K^+ X'$, so that a
state decaying 
as a \Bz (\Bzb) meson results most often in a \Kp (\Km). If
mixing takes place, the $\ell$ and the \kaon\ will have the same 
charge. Kaons produced in association with the $\ell\psoft$ pair are used to
measure the large instrumental asymmetry in kaon identification.

The observed asymmetry between the number 
of positive-charge and negative-charge leptons can be approximated as:
\begin{linenomath}
\begin{equation}\label{eq:arec}
A_\ell \simeq \Ar + \All \cdot \chi_d,
\end{equation}
\end{linenomath}
where $\chi_d=0.1862\pm0.0023$~\cite{PDG} is the integrated mixing 
probability for \Bz\
mesons, and \Ar\ is the charge asymmetry in the reconstruction of
\BtoDs\ decays. 

With the same approximations as before, 
the observed asymmetry in the rate of kaon-tagged mixed events is:
\begin{linenomath}
\begin{equation}
A_T = \frac{N(\ell^+\Kt^+) - N(\ell^-\Kt^-)} {N(\ell^+\Kt^+) + N(\ell^-\Kt^-)}
\simeq \Ar + \At + \All ,
\label{eq:amix}   
\end{equation}
\end{linenomath}
where \At\ is the charge asymmetry in kaon reconstruction.
A kaon with the same charge as the $\ell$
might also come from the CF decays  of the \Dz\
meson produced with the lepton from the partially reconstructed side (\Kr).
The asymmetry observed for these events is:
\begin{linenomath}
\begin{equation}
A_R =\frac{N(\ell^+\Kr^+) - N(\ell^-\Kr^-)} {N(\ell^+\Kr^+) + N(\ell^-\Kr^-)} \simeq \Ar+\At + \All \cdot \chi_d
\label{eq:asame} 
\end{equation}
\end{linenomath}
Equations \ref{eq:arec}, \ref{eq:amix}, and \ref{eq:asame}, { defining quantities computed in terms of the observed number of events integrated over time}, can be
inverted to extract \All\ and the detector induced asymmetries.
It is not possible to distinguish a \Kt from a 
\Kr in each event. They are separated on a statistical basis, using
kinematic features and proper-time difference information.

\section{ The \babar\ Detector}
\label{sec:det}

A detailed description of the \babar\ detector and the algorithms used
for charged and neutral particle reconstruction and identification is provided
elsewhere~\cite{babar_nim, babar_new}. A brief summary is given here.
The momentum of charged particles is measured by the tracking system, which
consists of a silicon vertex tracker (SVT) and a drift chamber (DCH) in a 1.5 T
magnetic field. The positions of points along the trajectories of charged particles
measured with the SVT are used for vertex reconstruction and for measuring
the momentum of charged particles, including those particles with low transverse
momentum that do not reach the DCH due to the bending in the magnetic field.
The energy loss in the SVT is used to discriminate low-momentum pions from 
electrons. 

Higher-energy electrons are identified from the ratio of the energy
of their associated shower in the electromagetic calorimeter (EMC) to their
momentum, the transverse profile of the shower, the energy loss in the DCH,
and the information from the Cherenkov detector (DIRC).
The electron identification efficiency is  $93\%$, and the  
misidentification rate for pions and kaons is less than $1\%$.

Muons are identified on the basis of the energy deposited in the EMC 
and the penetration in the instrumented flux return (IFR) of the 
superconducting coil, which contains resistive plate { chambers} and
limited stramer tubes interspersed with iron. Muon candidates 
compatible with the kaon hypothesis in the DIRC are rejected.
The muon identification efficiency is  about $80\%$, and the
misidentification rate for pions and kaons is $\sim 3\%$.

We select kaons from charged particles with momenta
larger than 0.2 \gevc\ using a standard algorithm which combines 
DIRC information with the
measurements of the energy losses in the SVT and DCH. True kaons are
identified with $\sim 85\%$ efficiency and a $\sim 3\%$ pion misidentification rate. 

\section{Event Selection}
\label{sec:evsel}

The data sample used in this analysis consists of 468 million \BB pairs, corresponding to
an integrated luminosity of 425.7\invfb, collected at the $\Y4S$ resonance (on-resonance) 
and 45\invfb collected 40\mev below the resonance (off-resonance) by the \babar\ detector~\cite{lumi}. 
The off-resonance events are used to describe the non-\BB\ (continuum) background.
A simulated sample of $\BB$ events with integrated luminosity equivalent to approximately 
three times the size of the data sample, based on EvtGen~\cite{evtgen} and GEANT4~\cite{geant4} with full detector response and event reconstruction, is used to test the analysis procedure.

We preselect a sample of hadronic events with at least four charged particles.
To reduce continuum background we require the ratio of the 2$^{nd}$ to the
0$^{th}$ order Fox-Wolfram variables~\cite{wolfram} be less than 0.6.
We then select a sample of partially reconstructed {\em B} mesons in the channel 
$\Bzb \rightarrow D^{*+} X \ell^{-} \bar{\nu}_{\ell}$, 
by retaining events containing a charged lepton ($\ell = e,\,\mu$) and a low-momentum 
pion (soft pion, $\pi^+_{s}$) from the decay $D^{*+}\to \Dz \pi^+_{s}$.
The lepton momentum must be in the range $1.4 < p_{\ellm} < 2.3 \gevc$ and 
the soft pion candidate must satisfy $60 < p_{\pi^{+}_{s}} < 190 \mevc$.
Throughout this paper the momentum, energy and direction of 
all particles are determined in the $e^+e^-$ rest frame.
The two tracks must be consistent with originating from a common vertex, constrained to the 
beam-spot in the plane transverse to the $e^+e^-$ collision axis. Finally, we
combine $p_{\ellm}$, $p_{\pi^{+}_{s}}$ 
and the probability from the vertex fit into a likelihood ratio
variable ($\eta$). A cut on  $\eta$ is
optimized to reject background from other \BB\ events. If more than one combination is
found in an event, we keep the one with the largest value of $\eta$.

The squared { missing mass} is:
\begin{linenomath}
\begin{equation}
{\mm} \equiv (E_{\mbox{\rm \small beam}}-E_{{D^*}} - 
E_{\ell})^2- ({\bf{p}}_{{D^*}}+{\bf{p}}_{\ell})^2 ,
\end{equation}
\end{linenomath}
where we neglect the momentum of the \Bz\ 
($p_B$ \mbox{$\approx$ 340 \mevc}) and identify the \Bz\ energy with the beam
energy $E_{\mbox{\rm \small beam}}$ in the $e^+e^-$ center-of-mass frame;
$E_{\ell}$ and  ${\bf{p}}_{\ell}$ are the energy and momentum of the
lepton and ${\bf{p}}_{{D^*}}$ is the estimated momentum of the $D^*$.
As a consequence of the limited phase space available in the $D^{*+}$
decay, the soft pion is emitted in a direction close to that of the 
$D^{*+}$ and a strong correlation
holds between the energy of the two particles in the \Bz center of mass frame. 
The $D^{*+}$ four-momentum can, therefore, be estimated by approximating 
its direction as that of the soft pion, and parameterizing its momentum as 
a linear function of the soft-pion momentum using simulated events.
We select pairs of tracks with opposite charge for the signal ($\ell^\mp \psoft^\pm$)
and we use same-charge pairs ($\ell^\pm \psoft^\pm$) for background studies.  

Several processes where $\dsp$ and $\ell^-$ originate from the same {\em B}-meson produce 
a peak near zero in the { \mm} distribution. The signal consists
of  (a) $\Bzb \rightarrow D^{*+} \ell^{-} \bar\nu_{\ell}$ decays (primary);
(b) $\Bzb \rightarrow D^{*+} (\mathrm{n}\pi) \ell^- \bar{\nu}_{\ell}$
(\dstrstr), and (c) $\Bzb \rightarrow D^{*+}\tau^- \bar{\nu}_{\tau} $, $\tau^- \rightarrow
\ell^{-}\bar{\nu}_{\ell}\nu_{\tau} $. The main source of peaking
background is due to charged-$B$ decays to resonant or non
resonant charm excitations, $\Bp \rightarrow D^{*+}
(\mathrm{n}\pi) \ell^- \bar{\nu}_{\ell}$, or to $\tau$ leptons, and  
 $B \rightarrow D^{*+} h^- X$, where the hadron ($h =
 \pi,K, D)$ is erroneously identified as, or decays to, a
 charged lepton.
 We also include radiative events, where photons with energy
above 1 MeV are emitted by any charged particle, as described in the simulation 
by PHOTOS~\cite{photos}. 
We define the signal region ${\mm} > -2~$GeV$^{2}/c^4$, and the sideband $-10 <  { \mm} < -4~$GeV$^{2}/c^4$.

\begin{center} 
\begin{figure}[htbp]
\includegraphics[width=9cm]{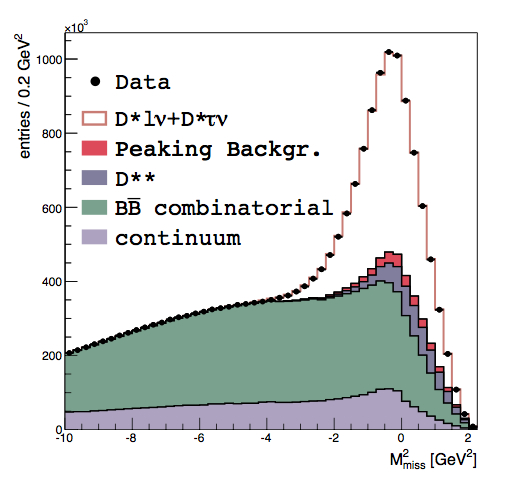}  
\caption{ (color online): {\mm} distribution for the selected
events. The data are represented by the points with error bars. The fitted contributions from
$\Bzb \rightarrow D^{*+} \ell^{-} \bar\nu_{\ell}$ plus $\Bzb \rightarrow D^{*+} \tau^{-} \bar\nu_{\ell}$, 
peaking background, $D^{**}$ events (1/3 from $\Bzb$ and 2/3 from $B^+$ decays),
  \BB\ combinatorial, and rescaled off-peak events are shown (see text for details).}
\label{f:fitmm}
\end{figure}
\end{center}

Continuum events and random combinations of a low-momentum 
pion and an opposite-charge lepton from
combinatorial \BB\ events contribute to the non-peaking background.
We determine the number of signal events in the sample with a minimum-$\chi^2$ fit to the {\mm} distribution in the 
interval $-10 <{\mm}< 2.5~$GeV$^2$/c$^4$. 
In the fit, the continuum contribution is obtained from off-peak
events, normalized by the on-peak to off-peak luminosity ratio, the
other contributions are taken from the simulation. The number of
events from combinatorial \BB\ background, primary decays and \dstrstr\ ((a) and (b) categories described previously) are allowed to vary in the
fit, while the other peaking contributions are fixed to the simulation
expectations (few percent).
The number of \Bz\
mesons in the sample is then obtained assuming that 2/3 of the fitted
{ number} of \dstrstr\ events are produced by \Bp\ decays, as suggested by simple
isospin considerations. 
We find a total of  $(5.945\pm 0.007) \cdot 10^6$ signal events, where the
uncertainty is only statistical. In the
full range signal events account for about 30\%\ of the sample and
continuum background for about 15\%.
The result of the fit is shown in
Fig.~\ref{f:fitmm}.

\section{Kaon Tag}
\label{sec:ktag}

{We indicate with \Kr (\Kt) kaons produced from the decay of the \Dz\ from the
partially reconstructed \Bz\ (\Brec), or in any step of the decay of the
other $B$ (\Btag).} 
We exploit the relation between the charge of the lepton and that of
the \Kt\ to define an event as ``mixed'' or ``unmixed''.
When an oscillation takes place, and the two \Bz\ mesons in the event have the same
flavor at decay time, a \Kt\ from a CF decay has the
same charge as the $\ell$. 
Equal-charge combinations are also
observed from Cabibbo-supressed (CS) \Kt production in unmixed events, and from CF \Kr\ production.
Unmixed CF \Kt, mixed CS \Kt, and CS \Kr, result in opposite-charge
combinations. 
Other charged particles wrongly identified as kaons contribute both to 
equal and opposite charge events with comparable rates. 

We distinguish \Kt\  from \Kr\ using proper-time difference information.
We define $\dZ = \Zr - \Zt$,
where \Zr\ is the projection along the beam direction of the \Brec\ decay point, 
and \Zt\ is the projection along the same direction of the 
intersection of the \kaon\ track trajectory with the beam-spot.
In the boost approximation~\cite{BAPX} we measure the
proper-time difference between the two {\em B} {meson decays} using the relation
$\DT = \dZ /( \beta \gamma c)$, where the parameters $\beta,
\gamma$ express the Lorentz boost from the laboratory to the \FourS\ rest frame.
We reject events if the uncertainty $\sigma(\DT)$ exceeds 3 ps.

Due to the short lifetime and small boost of the \Dz\ meson, 
small values of \DT\ are expected for the \Kr. 
Much larger values are instead expected for CF mixed \Kt,
due to the long period of the \Bz\ oscillation.  Fig.~\ref{f:dz} shows the \DT\ 
distributions for \Kt and \Kr events,
as obtained from the simulation.
To improve the separation between \Kt\ and \Kr, we also exploit kinematics. 
In the rest frame of the \Bz, the $\ell$ and the \dsp\ are emitted
at large angles. Therefore the angle $\theta_{\ell\kaon}$
between the $\ell$ and the \Kr\ has values close to $\pi$, and
$\cos\theta_{\ell\kaon}$ close to -1. The  corresponding distribution
for \Kt\ is instead uniform, as shown in Fig.~\ref{f:costhe-mc}.

In about 20\% of the cases, our events contain more than one kaon:
most often we find both a \Kt\ and a \Kr\ candidate. 
As these two carry different information, we accept multiple candidate
events. Using several simulated pseudo-experiments, we assess the effect of this choice on the statistical uncertainty.

\section{ Extraction of \dCP}
\label{sec:fit}
The measurement proceeds in two stages. 

First we measure the sample
composition of the eight
tagged samples grouped by lepton kind, lepton charge and $K$
charge, with the fit to {\mm} previously described. We also fit the four
inclusive lepton samples to determine the charge asymmetries at the
reconstruction stage (see Eq.~\ref{eq:arec}). { At this point of the analysis we use the total number of collected events.}

The results of the first stage are 
used
in the second
stage, where we fit simultaneously the $\cos\theta_{\ell K}$ and \DT\
distributions in the eight tagged samples.

The \DT\ distributions for $B \Bbar$, $BB$ and $\Bbar \Bbar$ events are
parameterized as the convolutions of the theoretical distributions 
${\cal F}_i(\dT|\vec{\Theta})$ with
the resolution function ${\cal R} (\DT,\dT)$: 
${\cal G}_i(\DT) = \int_{-\infty}^{+\infty}
{\cal F}_i(\dT|\vec{\Theta}) {\cal R} (\DT,\dT) d(\dT)$, where $\dT$
is the actual difference between the times of decay of the two mesons and
$\vec{\Theta}$ is the vector of the physical parameters.
The decays of the \Bp\ mesons are parameterized by an exponential function, 
 \mbox{${\cal F}_{\Bp} =\Gamma_+ e^{-|\Gamma_+\dT|}$}, where  the \Bp\ decay
width is the inverse of the lifetime 
{$\Gamma_+^{-1} = \tau_+ = 1.641\pm 0.008$ ps~\cite{PDG}}.
According {to} Ref.~\cite{DCS}, the decays of the \Bz\ mesons are 
described by the following expressions:

\begin{widetext} 
{\small
\begin{eqnarray}
\label{eqn:PDF1}
\mathcal{F}_{\Bzb \Bz}(\dT) &=&
{\cal E}(\dT) 
\bigg[
\left(1+\bigg\vert \frac{q}{p}\bigg\vert^{2} r'^2 \right)\cosh(\Delta\Gamma \dT/2) +
 \left(1-\bigg\vert \frac{q}{p}\bigg\vert^{2} r'^2 \right)\cos(\Delta m \dT) - 
 \bigg\vert{\frac{q}{p}}\bigg\vert  (b+c) \sin(\Delta m \dT)\bigg]  \\
\label{eqn:PDF2}
\mathcal{F}_{\Bz\Bzb}(\dT) &=& 
{\cal E}(\dT) 
\bigg[
\left(1+\bigg\vert \frac{p}{q}\bigg\vert^{2} r'^2 \right)\cosh(\Delta\Gamma \dT/2) + 
 \left(1-\bigg\vert \frac{p}{q}\bigg\vert^{2} r'^2 \right)\cos(\Delta m \dT) + 
 \bigg\vert{\frac{p}{q}}\bigg\vert  (b-c) \sin(\Delta m \dT)\bigg] \\
\label{eqn:PDF3}
\mathcal{F}_{\Bzb\Bzb}(\dT) &=& 
{\cal E}(\dT) 
\bigg[
\left(1+\bigg\vert \frac{p}{q}\bigg\vert^{2} r'^2 \right)\cosh(\Delta\Gamma \dT/2) - 
 \left(1-\bigg\vert \frac{p}{q}\bigg\vert^{2} r'^2 \right)\cos(\Delta m \dT) - 
 \bigg\vert{\frac{p}{q}}\bigg\vert  (b-c) \sin(\Delta m \dT)\bigg]  
\bigg\vert \frac{q}{p} \bigg \vert^{2} \\
\label{eqn:PDF4}
\mathcal{F}_{\Bz\Bz}(\dT) &=& 
{\cal E}(\dT) 
\bigg[
\left(1+\bigg\vert \frac{q}{p}\bigg\vert^{2} r'^2 \right)\cosh(\Delta\Gamma \dT/2) - 
 \left(1-\bigg\vert \frac{q}{p}\bigg\vert^{2} r'^2 \right)\cos(\Delta m \dT) + 
 \bigg\vert{\frac{q}{p}}\bigg\vert  (b+c) \sin(\Delta m \dT)\bigg]
\bigg\vert
\frac{p}{q} \bigg \vert^{2} 
\\ \nonumber 
{\cal E}(\dT) &=& \frac{\Gamma_0}{2(1+r'^2)} e^{-\Gamma_0|\dT|},
\end{eqnarray}
}
\end{widetext}
where the first index refers to the flavor of the  \Brec\ and the second 
index to that of the \Btag.
In Eqs.~\ref{eqn:PDF1}-\ref{eqn:PDF4}, $\Gamma_0 = \tBz^{-1}$ is the average width of the two \Bz\
mass eigenstates, $\Delta m$ and $\Delta \Gamma$ are respectively their mass
and width differences, $r'$ is a 
parameter resulting from the interference of CF and doubly Cabibbo-suppressed (DCS) decays on the \Btag\ side, $b$ and $c$ two parameters
expressing the \CP violation arising from that interference. In the
Standard Model the value of $r'$ is rather small, ${\cal O} ( 1\%)$, \mbox{$b = 2r' \sin(2\beta+\gamma) \cos\delta'$}, and 
 \mbox{$c = -2r' \cos(2\beta+\gamma) \sin\delta'$}, where $\beta$ and
 $\gamma$ are angles of the Unitary Triangle~\cite{UT}, and $\delta'$
 is a strong phase. 
Besides $|q/p|$, also $\Delta m$, \tBz , $\sin(2\beta+\gamma)$, $b$, and $c$ are determined as
effective parameters to reduce the systematic uncertainty.
The value of $\Delta \Gamma$ is fixed to zero, and then varied
within its allowed range~\cite{PDG} when computing the systematic uncertainty.
Neglecting the tiny contribution from DCS
decays, the main contribution to the asymmetry is time independent and due to the normalization factors in Eqs. \ref{eqn:PDF3} and \ref{eqn:PDF4}.

When the \Kt\ comes from the decay of the \Bz\ meson to a
\CP-eigenstate (as, for instance $\Bz \rightarrow D^{(*)}D^{(*)}$), a
different expression applies:
\begin{linenomath}
\begin{eqnarray}
\mathcal{F}_{CP}(\dT) = &\frac{\Gamma_0}{4}&
e^{-\Gamma_0|\dT|} [ 1\pm S \sin(\Delta m \dT)\\
\nonumber
 &\pm& C \cos(\Delta m \dT )],
\end{eqnarray}
\end{linenomath}
where the sign plus is used if the \Brec\ decays as a \Bz and the sign
minus otherwise. {This sample contains several components and is strongly biased by the selection cuts,
therefore} we take the values of $S$ and $C$, and the fraction of
these events in each sample (about 1\%) from the simulation.

The resolution function ${\cal R} (\DT,\dT)$ accounts for the
uncertainties in the measurement of \DT, 
for the effect of the boost approximation,
and for the displacement of the \Kt\
production point from the \Btag\ decay position due to the motion of the
charm meson. It consists of the superposition of several Gaussian
functions convolved with exponentials.

We determine $\dCP$ with two different inputs for ${\cal G}_{\Kr}(\DT)$, the \DT\  distribution for \Kr events, and take the mean value of the two determinations as the nominal result. As first input, we use the distribution obtained from a high purity selection of \Kr events on data, ${\cal G}^{\rm Data}_{\Kr}(\DT)_{HP}$. As second input, we use the distribution for \Kr events as predicted by the simulation, ${\cal G}^{\rm MC}_{\Kr}(\DT)$, corrected using Eq.~\ref{eq:dtagcor}.
\begin{linenomath}
\begin{equation}
{\cal G}^{\rm Data}_{\Kr}(\DT)={\cal G}^{\rm MC}_{\Kr}(\DT)\times \left(\frac{{\cal G}^{\rm Data}_{\Kr}(\DT)}{{\cal G}^{\rm MC}_{\Kr}(\DT)}\right)_{HP}
\label{eq:dtagcor}
\end{equation}
\end{linenomath}
To select the high purity \Kr sample, we require the  
lepton and the kaon to have the same charge. 
As discussed above, this sample consists of about 75\% genuine \Kr, where the 
residual {number} of events with a \Kt\ is mostly due to mixing. 
Therefore we select events with a second
high-momentum lepton, with charge opposite to that of the
first lepton. According to the simulation, this raises the \Kr\ purity
in the sample to about 87\%. Finally, we use topological variables,
correlating the kaon momentum-vector to those of the two
leptons, to raise the \Kr\ purity in the sample to about 95\%.

Due to the large number of events, the fit complexity, and the high number of floated parameters, the time needed
for an unbinned fit to reach convergence is too large, therefore we perform a binned maximum likelihood fit. Events
belonging to each of the {eight} categories are
grouped into 100 \DT\ bins, 25 $\sigma(\DT)$ bins, 4 cos$\theta_{\ell,K}$
bins, and 5 {\mm} bins. We further split { the} data into five bins of $K$
momentum, $p_K$, to account for the dependencies of several parameters,
describing the \DT\ resolution
function, the cos$(\theta_{\ell K})$ distributions, the fractions
of \Kt\ events, etc.,  observed in the simulation. 

Accounting for events with wrong flavor assignment 
and \Kr\ events, the peaking \Bz\
contributions 
to the equal and opposite charge samples in each bin $j$ are:
{
\begin{widetext}
\begin{linenomath}
{\small
\begin{eqnarray}
\label{eq:sigpdf}
{{\cal G}^{\Bz}_{\ell^+ K^+} (j)}&=& (1+\Ar)(1+\At)
 \{(1-f_{\Kr}^{++}) [(1-\omega^+) {\cal
  G}_{\Bz\Bz} (j)+\omega^- {\cal G}_{\Bz {\Bzb}}(j) ] +
 f_{\Kr}^{++}(1-\omega'^{+}){\cal G}_{\Kr}(j) (1+\chi_d \All)\} \\
\nonumber
{{\cal G}^{\Bz}_{\ell^- K^-} (j)}&=& (1-\Ar)(1-\At)  \{(1-f_{\Kr}^{--}) [(1-\omega^-) {\cal
  G}_{{\Bzb} {\Bzb}} (j)+\omega^+ {\cal G}_{{\Bzb} \Bz}(j) ]
+ f_{\Kr}^{--} (1-\omega'^{-}){\cal G}_{\Kr}(j) (1-\chi_d \All)\}\\
\nonumber 
{{\cal G}^{\Bz}_{\ell^+ K^-} (j)}&=& (1+\Ar)(1-\At)
 \{(1-f_{\Kr}^{+-}) [(1-\omega^-) {\cal
  G}_{\Bz\Bzb} (j)+\omega^+ {\cal G}_{\Bz \Bz}(j) ] +
 f_{\Kr}^{+-}{\omega'^{+}}{\cal G}_{\Kr}(j) (1+\chi_d \All)\} \\
{{\cal G}^{\Bz}_{\ell^- K^+} (j)}&=& (1-\Ar)(1+\At)
 \{(1-f_{\Kr}^{-+}) [(1-\omega^+) {\cal
  G}_{\Bzb\Bz} (j)+\omega^- {\cal G}_{\Bzb \Bzb}(j) ] +
 f_{\Kr}^{-+}{\omega'^{-}}{\cal G}_{\Kr}(j) (1-\chi_d \All)\} 
\nonumber
\end{eqnarray}
}
\end{linenomath}
\end{widetext}
{where the probability density functions (PDFs) ${\cal G}_{\Bz\Bz}(\DT)$, ${\cal G}_{\Bz {\Bzb}}(\DT)$, ${\cal G}_{{\Bzb} \Bz}(\DT)$ and ${\cal G}_{{\Bzb} {\Bzb}}(\DT)$ are the convolutions of the theoretical distributions in Eqs.~\ref{eqn:PDF1}-\ref{eqn:PDF4} with the resolution function.}
The reconstruction asymmetries $\Ar$ are determined separately for the
$e$ and $\mu$ samples.  
Wrong-flavor assignments are described by the probabilities $\omega^\pm$ for \Btag~and $\omega'^\pm$ for \Brec.
They are different because \Kt\ come from a mixture of $D$ mesons, while \Kr\
are produced by \Dz\ decays only.
The parameters $f_{\Kr}^{\pm \pm}(p_K)$ describe the
fractions of \Kr\ tags {in each} sample as a function of the kaon momentum. 
Due to the different charge asymmetry of the \Kt\ and the \Kr\ events, 
(see equations~\ref{eq:amix} and \ref{eq:asame}), 
the fitted values of $f_{\Kr}^{\pm \pm}({p_K})$ and $|q/p|$ are strongly correlated.
The $f_{\Kr}^{\pm \pm}({p_K})$ fractions can be factorized as:
\begin{linenomath}
\begin{equation}
f_{\Kr}^{\pm \pm} (|q/p|)=f_{\Kr}^{\pm \pm}(|q/p|=1)\times g^{\pm \pm}(|q/p|)
\end{equation}
\end{linenomath}
where the $f_{\Kr}^{\pm \pm}(|q/p|=1)$ parameters are left free in the fit and 
$g^{\pm \pm}(|q/p|)$ 
are analytical functions.
In order to limit the number of free parameters in the fit, the fractions 
of \Kr\ events in the \Bp\ sample
are computed from the corresponding fractions in the \Bz\ samples:
\begin{linenomath}
\begin{equation}
f^{\pm \pm}_{\Kr}(\Bp)=f^{\pm \pm}_{\Kr} (|q/p|=1)\times R^{\pm \pm}
\label{eq:rmc}
\end{equation}
\end{linenomath}
where $R^{\pm \pm}$ are correction factors
obtained from the simulation.

The combinatorial background consists of \Bp\ and \Bzb\ decays with comparable
contributions.
A non-negligible fraction of \Bzb\ combinatorial events is obtained
when the lepton in $B\to D^*X\ell\nu$ decay is combined with a soft pion from the decay of a tag-side
\dsp . As the two particles must have opposite charges, the fraction
of mixed events in the \Bzb\ combinatorial background is larger than 
for peaking events. In the simulation we find that the effective
mixing rate of the combinatorial events depends linearly on the kaon momentum according to the
relation:
\begin{linenomath}
\begin{equation}
 \chi^{\rm comb}_{d} = \chi^{\rm comb}_{0} ( a + b \cdot  p_K),
\label{eq:effchi}
\end{equation} 
where
\begin{equation}
\chi^{\rm comb}_{0} = \frac{x_{\rm comb}^2 }{2(1+x_{\rm comb}^2)}  
\end{equation}
\end{linenomath}
and
$x_{\rm comb} = \Delta m^{\rm comb} \tbz^{\rm comb}$. 
In this expression, $\Delta m^{\rm comb}$ and  $\tbz^{\rm comb}$
are the mass difference and lifetime measured in combinatorial
events. 
To account for this effect, we
use for \Bzb\ combinatorial background the same expressions as for the
signal (see Eq.~\ref{eq:sigpdf}), with the replacements: 
\begin{linenomath}
\begin{equation}
{\cal G}^{\rm comb}_{\Bzb \Bzb} ={\cal  G}_{\Bzb \Bzb } \frac {\chi^{\rm comb}_d }{\chi^{\rm comb}_0 },
\end{equation}
$${\cal G}^{\rm comb}_{\Bz \Bz} = {\cal  G}_{\Bz \Bz }\frac{ \chi^{\rm comb}_d }{\chi^{\rm comb}_0},$$
$${\cal  G}^{\rm comb}_{\Bzb \Bz} = {\cal  G}_{\Bzb \Bz }\frac{1-\chi^{\rm comb}_d}{1-\chi^{\rm comb}_0 }
,$$
$${\cal  G}^{\rm comb}_{\Bz \Bzb} = {\cal  G}_{\Bz \Bzb } \frac{1-\chi^{\rm comb}_d }{1-\chi^{\rm comb}_0 }
.$$
\end{linenomath}
The parameters $a$ and $b$ in Eq.~\ref{eq:effchi}, $\Delta m^{\rm comb}$ and  $\tbz^{\rm comb}$
are determined in the fit.

The probabilities to assign a wrong flavor to \Btag\ 
in the combinatorial sample are found to be different in mixed and unmixed events. 

Different sets of parameters are used for peaking and for combinatorial
events, including lifetimes, frequencies
of \Bzb\ oscillation, detector related asymmetries, whereas the same value of 
 \magqp\ is used.
For \Bp\ combinatorial events, the same PDFs as for peaking \Bp\ background are employed, with different
sets of parameters.

The distribution ${\cal G}_{\rm cont} (\DT)$ of continuum events is represented by a decaying 
exponential, convolved
with a resolution function similar to that used for { $B$ events}. The
effective lifetime and resolution parameters are determined 
by fitting simultaneously the
off-peak data.

We rely on the simulation to parameterize the $\cos\theta_{\ell K}$ distributions.
The individual $\cos\theta_{\ell K}$ shapes for the eight \BB\ tagged samples are obtained 
from the histograms of the corresponding
simulated distributions, separately for \Kt\ and \Kr\ events, whereas we interpolate off-peak data to describe the continuum.

{The normalized $\DT$ distributions for each tagged sample are then
expressed as the sum of the predicted contributions from peaking, 
\BB\ combinatorial, and continuum background events:}

\begin{widetext}
\begin{linenomath}
{\small
\begin{eqnarray} 
\nonumber
&&{{\cal F}^{\ell\kaon}(\DT, \sigma_{\DT}, \mm, \mathrm{cos}\theta_{\ell,K}, p_K | \tBz, \Delta m, |q/p|)} =\\
\nonumber
&&(1-f_{\Bp}({\mm})-f_{CP}(\mm)-f_{\rm comb}(\mm)-f_{\rm cont}(\mm)){{\cal G}^{\Bz}_{\ell\kaon}(\DT, \sigma_{\DT}, \mathrm{cos}\theta_{\ell,K}, p_K )}\\
&&+ f_{\Bp}(\mm) {{\cal G}^{\Bp}_{\ell\kaon}(\DT, \sigma_{\DT}, \mathrm{cos}\theta_{\ell,K}, p_K)}
\nonumber
+ f_{CP}(\mm) {{\cal G}^{CP}_{\ell\kaon}(\DT, \sigma_{\DT}, \mathrm{cos}\theta_{\ell,K}, p_K)}
+ f^0_{\rm comb}(\mm) {{\cal G}^{\Bz \rm comb}_{\ell\kaon}(\DT, \sigma_{\DT}, \mathrm{cos}\theta_{\ell,K}, p_K)}\\
&&+ f^+_{\rm comb}(\mm) {{\cal G}^{\Bp \rm comb}_{\ell\kaon}(\DT, \sigma_{\DT}, \mathrm{cos}\theta_{\ell,K}, p_K)}
+ f_{\rm cont}(\mm) {{\cal G}^{\rm cont}_{\ell\kaon}(\DT, \sigma_{\DT}, \mathrm{cos}\theta_{\ell,K}, p_K)} 
\end{eqnarray}
}
\end{linenomath}
\end{widetext} 
where the fractions of peaking \Bp\ ($f_{\Bp}$), \CP eigenstates
($f_{CP}$), combinatorial \BB\ ($f_{\rm comb}$), and continuum ($f_{\rm cont}$)
events in each \mm\ interval are taken from the results of the first
stage. The fraction of \Bz ($f^0_{\rm comb}$) and of \Bp\ events ($f_{\rm comb}^+
= f_{\rm comb}- f^0_{\rm comb}$) in the combinatorial background have been determined from
a simulation. { The functions ${\cal G}^{\Bz}_{\ell\kaon}(j)$ for peaking \Bz events are defined in Eq.~\ref{eq:sigpdf}; the functions
${\cal G}^{\Bp}_{\ell\kaon}(j)$, ${\cal G}^{CP}_{\ell\kaon}(j)$, ${\cal G}^{\Bz \rm comb}_{\ell\kaon}(j)$, ${\cal G}^{\Bp \rm comb}_{\ell\kaon}(j)$ and ${\cal G}^{\rm cont}_{\ell\kaon}(j)$ are the corresponding PDFs for the other samples.}

For a sample of \Bz\ signal events tagged by a kaon from the \Btag ~meson decay, 
the expected fraction $P_{m}^{\rm exp}$ of mixed events in each $p_K$ bin depends on $\Delta m$, $\tbz$, and $\omega^{\pm}$, and reads
\begin{equation}
P_{m}^{\rm exp}=\frac{{{\cal G}^{\Bz_T}_{\ell^+ K^+}}+{{\cal G}^{\Bz_T}_{\ell^- K^-}}}{{{\cal G}^{\Bz_T}_{\ell^+ K^+}}+{{\cal G}^{\Bz_T}_{\ell^- K^-}}+{{\cal G}^{\Bz_T}_{\ell^+ K^-}}+{{\cal G}^{\Bz_T}_{\ell^- K^+}}},
\end{equation}
{ where the functions ${\cal G}^{\Bz_T}_{\ell K}$ are obtained from Eq.~\ref{eq:sigpdf} taking into account only the contributions from $\Bz_T$ events.}

We estimate this fraction by
multiplying the likelihood by the binomial factor
\begin{equation}
{C^{B^0_T}_{m}}=\frac{N!}{N_{m}! N_{u}!}(P_{m}^{\rm exp})^{N_{m}}(1-P_{m}^{\rm exp})^{N_{u}},
\end{equation}
where $N_m$ and $N_u$ are the number of mixed and unmixed events, {respectively, in a given $p_K$ bin for each subsample. These are obtained as the sums of the numbers of mixed events tagged by a kaon of a given charge, 
\begin{align}
N_{m} = N_{m,K^+} + N_{m, K^-}\\ 
N_{u} = N_{u,K^+} + N_{u, K^-}.
\end{align} 
Finally, $N=N_m+N_u$.}

The corresponding value of  $P_{m,\rm comb}^{\rm exp}$
for the sample of \Bz\ combinatorial events tagged by a kaon from the \Btag~meson decay
{ depends on $\Delta m^{\rm comb}$, $\tbz^{\rm comb}$}, and the
wrong flavor assignment probability for the mixed and unmixed subsamples.

For a sample of \Bz\ events tagged by a kaon from the \Btag~meson decay, 
the expected fraction of mixed  {and} unmixed events, tagged by a positive charge kaon, depends on \All and the detector charge asymmetries. For the \Bz\ signal sample this fraction  
reads
\begin{equation}
P_{m(u), K^+}^{\rm exp}=\frac{{\cal G}^{B^{0}_T}_{\ell^+(\ell^-) K^+}}{{\cal G}^{B^{0}_T}_{\ell^+(\ell^-) K^+}+{\cal G}^{B^{0}_T}_{\ell^-(\ell^+) K^-}}
\end{equation}

We estimate this { quantity} by
multiplying the likelihood by the binomial factor
\begin{eqnarray}
{C^{B^0_T}_{m(u),K^+}}&=&\frac{N_{m(u)}!}{N_{m(u),K^+}! N_{m(u),K^-}!}\\ \
\nonumber
&&{(P_{m(u) K^+}^{\rm exp})}^{N_{m(u) K^+}}\\
&&(1-P_{m(u) K^+}^{\rm exp})^{N_{m(u) K^-}}.
\nonumber
\end{eqnarray}

{
For a sample of \Bz\ events tagged by a kaon from the \Brec~meson decay, 
the fraction of mixed events depends on $\omega'^{\pm}$.
The fraction of mixed  and unmixed events, tagged by a positive charge kaon, depends on the detector charge asymmetries and on \All.
Analogously, the corresponding fractions for a sample of \Bp\ events tagged by a kaon from the \Btag\ or \Brec\ meson decay, give information on the detector charge asymmetries.
}
 
The same values of \All and \At are shared between signal and combinatorial \Bz\ samples.
The values of $P_{m}^{\rm exp}$ and $P_{m(u), K^+}^{\rm exp}$ for all the subsamples are obtained from the ratio of integrals of the corresponding observed PDFs.
{
We maximize the likelihood
\begin{widetext}
\begin{eqnarray}
\nonumber
{\cal L}= \left( \prod_{i=1}^{N_{m, K^+}}{\cal F}^{\ell^+ K^+}_i \right) 
\left( \prod_{j=1}^{N_{m, K^-}}{\cal F}^{\ell^- K^-}_j \right)
\left( \prod_{m=1}^{N_{u, K^+}}{\cal F}^{\ell^- K^+}_m \right)\left( \prod_{n=1}^{N_{u, K^-}}{\cal F}^{\ell^+ K^-}_n \right)\times \left(\prod_{k=1}^{5}\prod_{l=1}^{8} C^l_m(k) C^l_{m, K^+}(k) C^l_{u, K^+}(k)  \right)
\end{eqnarray}
\end{widetext}
where the indices $i$, $j$, $m$ and $n$ denote the mixed (unmixed) events, tagged by a kaon of a given charge, the index $k$ denotes the $p_K$ bin, and the index $l$ denotes the signal (combinatorial background) subsample, according to the $B$ meson charge $(\Bz$ or $\Bp)$, and the tagging kaon category ($K_T$ or $K_R$).
}

A total of 168 parameters are determined in the fit. To reach the convergence of the fit,
we use a three step approach. In the first step we fit only the parameters describing the \Brec\ event fractions,
whereas all the other parameters are fixed to the values obtained on simulated events.
In the second step we fix the \Brec\ fractions to the values obtained in the first step and we float only the parameters describing the resolution function. In the last
step we fix the resolution parameters to the values obtained in the second step and we float again all the other parameters together with $\dCP$.
\begin{center} 
\begin{figure}[htbp] 
\includegraphics[width=9cm]{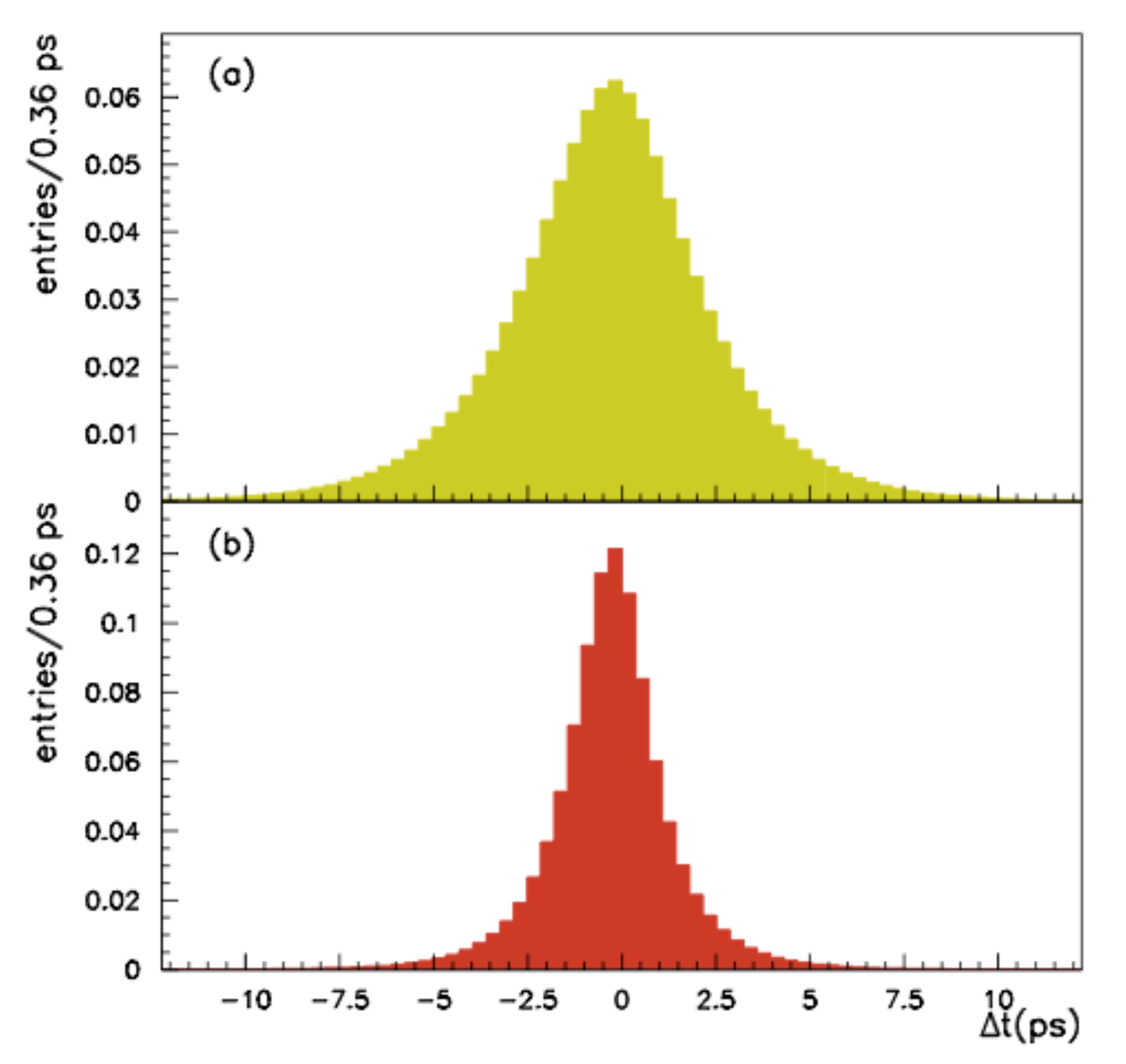}  
\caption{(color online): \DT\ distributions for \Kt (a)
  and for \Kr (b), as predicted by the simulation.}
\label{f:dz}
\end{figure}
\end{center}
\begin{center} 
\begin{figure}[htbp] 
\includegraphics[width=9cm]{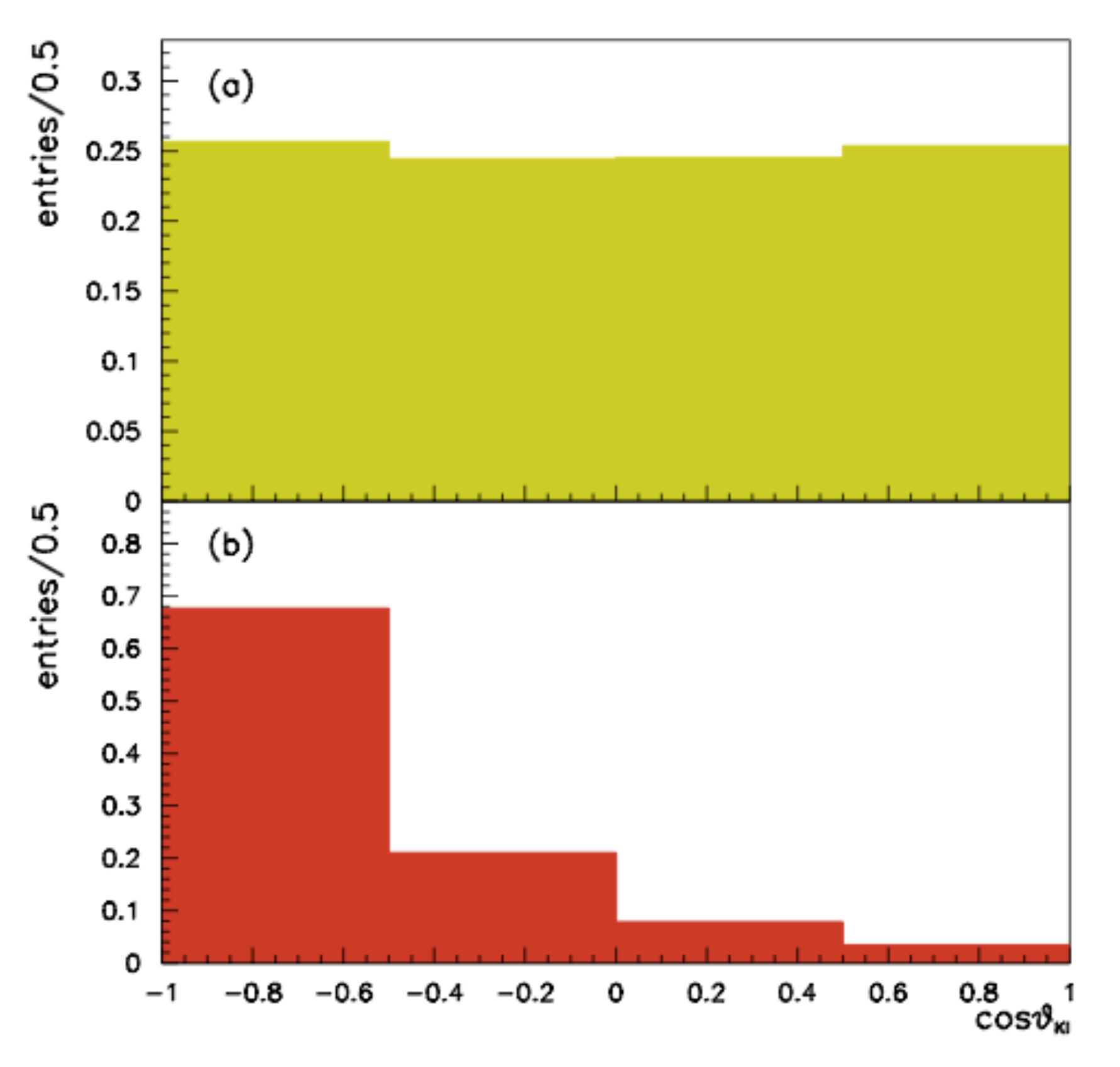}
\caption{cos$(\theta_{\ell\kaon})$ distributions for \Kt (a)
  and for \Kr (b), as predicted by the simulation.}
\label{f:costhe-mc}
\end{figure}
\end{center}
\section{Fit Validation}
\label{sec:val}

Several tests are performed to validate the result.
We analyze simulated events with the same procedure we use for data, first considering
  only \Bz\ signal and adding step by step all the other 
  samples. At each stage, the fit reproduces the generated values of $\dCP$ (zero),
  and of the other most significant parameters (\Ar, \At, $\Delta m$, and
  \tBz). 

We then repeat the test, randomly rejecting \Bz or $\Bzb$ events in order to produce
  samples of simulated events with $\dCP = \pm0.005, \pm0.01, \pm0.025$.
 Also in  this case the generated values are well reproduced by the fit.
By removing events we also vary artificially \Ar\ or \At , testing
values in the range of $\pm 10\%$. In each case, the input values are
correctly determined, and an unbiased value of $\dCP$ is always 
obtained. A total of 67 different simulated event samples are used
to check for biases.

Pseudo experiments are used to check the result and its statistical uncertainty. We perform
  173 pseudo-experiments, each with the same number of events
  as the data. 
We obtain a value of the likelihood larger than in data  
in $23\%$ of the cases.

The distribution of the fit results for $\dCP$, {obtained using the MIGRAD minimizer of the MINUIT~\cite{minuit} physics analysis tool for function minimization},
 is described by a Gaussian function with a central value biased by 
$-3.6\times 10 ^{-4}$ (0.4 $\sigma$) with respect to the nominal result.
We quote this discrepancy as a systematic uncertainty related to the analysis bias.

The pull distribution is described by a Gaussian
  function, with a central value  $-0.48\pm0.11$
  and RMS width of $ 1.44 \pm 0.08$. 
The statistical uncertainty, is, therefore, somewhat underestimated.
{ As a cross check,} by fitting the negative log likelihood profile near the minimum with 
a parabola, we obtain an estimate of the statistical uncertainty { 
from the $\dCP$ values for which $-\log {\cal L} = -\log {\cal L}_\mathrm{min} + 0.5$~\cite{PDG}. This result is} in good agreement 
with the RMS width of the distribution of the pseudo-experiments results, { which we take as the 
statistical uncertainty of the measurement.}

\section{Systematic Uncertainties and Consistency Checks}
\label{sec:syst}

We consider several sources of systematic uncertainty. We vary each
quantity by its error, as discussed below, we repeat the
measurement, and we consider the 
variation of the result as the corresponding systematic uncertainty.
We then add in quadrature all the contributions to determine the overall
systematic uncertainty.

{\it Peaking Sample Composition:} we vary the sample composition in the second-stage fit by the statistical uncertainties obtained at the first
stage; the corresponding variation is added in
quadratuture to the systematic uncertainty. 
We then vary the fraction of \Bz\ to \Bp\ in the \dstrstr peaking sample
in the range {$(50\pm25) \%$} to account for (large) violation of isospin symmetry.
The fraction of the peaking contributions fixed to the simulation expectations 
is varied by $\pm20\%$.
Finally we conservatively vary the
fraction of \CP-eigenstates by $\pm 50\% $.

{\it \BB\ combinatorial sample composition:} the fraction of \Bp\ and \Bz\ 
in the \BB\ combinatorial background is determined by the simulation. 
A difference between \Bp\ and \Bz\ is expected when mixing takes
place and the lepton is coupled to the tag side \psoft\ from 
${\Bzb} \rightarrow D^{*+}  X$ decay.
We then vary the fraction of \Bz to \Bp\ events in the combinatorial sample
by $\pm 4.5\%$, which corresponds to the uncertainty in the
inclusive branching fraction ${\Bzb} \rightarrow \dsp X$.

{\it \DT\ resolution model:}  
in order to reduce the time in the fit validation, all the 
parameters describing the 
resolution function, which show a weak correlation with $|q/p|$, are fixed 
to the values obtained using an iterative procedure. 
We perform a fit by leaving free all the parameters and we quote
the difference between the two results as a systematic uncertainty.

{\it \Kr\ fraction:} we vary the fraction of $\Bp \rightarrow \Kr X$ to
$\Bz \rightarrow \Kr X$ by $\pm 6.8\%$, which corresponds to
the uncertainty on the ratio 
$BR(D^{*0}\rightarrow K^-X) / BR(D^{*+}\rightarrow K^-X)$.

{\it \Kr\ \DT\ distribution:} 
we use half the difference between the results obtained using the two 
different strategies to describe the \Kr\ \DT\ distribution as a systematic
uncertainty.

{\it Fit bias:} 
we consider two contributions:
the statistical uncertainty
on the validation test using the detailed simulation, and the difference between the
nominal result and the central result determined from the ensemble of parameterized
simulations, described in {Sec.~\ref{sec:val}}.

{\it \CP eigenstates description:}
we vary the $S$ and $C$ parameters describing the {\CP eigenstates} by their
statistical uncertainty as obtained from simulation.

{\it Physical parameters:}
we repeat the fit setting the value of $\Delta\Gamma$ to 0.02 ps$^{-1}$ instead of zero.
The lifetime of the $B^0$ and $B^+$ mesons and $\Delta m$
are floated in the fit. Alternatively, we check the effect of fixing each parameter
in turn to the world average.

By adding in quadrature all the contributions described above, and
summarized in Table  \ref{t:syserr}, we
determine an overall systematic uncertainty of ${{+1.88}\atop{-1.61}} \times 10^{-3}$. 
\begin{center}
\begin{table}
\caption{Breakdown of the main systematic uncertainties {on \dCP}.}
\label{t:syserr}
\begin{center}
\begin{tabular}{l|c}
\hline \hline
Source & $\delta\dCP ~[10^{-3}]$\\
\hline
\raisebox{0mm}[3mm][0mm]{Peaking} sample composition & ${+1.50}\atop{-1.17}$\\
Combinatorial sample composition & $\pm0.39$\\ 
\DT\ resolution model & $\pm 0.60$\\ 
\Kr\ fraction & $\pm0.11$\\ 
\Kr\ \DT\ distribution & $\pm 0.65$\\ 
Fit bias & ${+0.58}\atop{-0.46}$\\ 
\CP eigenstate description & $-$\\ 
Physical parameters & ${+0}\atop{-0.28}$\\ 
\hline
\raisebox{0mm}[3mm][0mm] Total & ${+1.88}\atop{-1.61}$
\end{tabular}
\end{center}
\end{table}
\end{center}

\section{Results}
\label{sec:res}

We perform a blind analysis: the value of $\dCP$ is kept masked
until the study of the systematic uncertainties is completed and all
the consistency checks are succesfully accomplished; the values of all
the other fit parameters are not masked.

After unblinding we find:
$\dCP=(0.29\pm0.84)\times 10^{-3}$.
We report in Table \ref{t:res} the fit
results for the most significant parameters.
\begin{table*}
\caption{Fit results for the most significant parameters with their statistical uncertainty. 
Second column: fit to the data; third: fit to
  simulated events; last: values of the parameters in the simulation
  at generation stage. The detector asymmetries in the last column are obtained
from the comparison of the reconstruction efficiencies for positive and negative particles in the simulation.}
\label{t:res}
\begin{center}
\begin{tabular}{l|c|c|c}
\hline \hline
Parameter & Data {fit} & Simulation {fit}& From MC information \\
\hline
$\dCP$   &$0.0003\pm0.0008$ &$0.0003\pm0.0005$ & 0 \\
\Are  &$0.0030\pm0.0004$ &$0.0097\pm0.0002$ &$0.0090\pm0.0003$  \\
\Arm  &$0.0031\pm0.0005$ &$0.0084\pm0.0003$ &$0.0091\pm0.0003$ \\
\At   &$0.0137\pm0.0003$ &$0.0147\pm0.0001$ & $0.0151\pm0.0001$ \\
\tBz~(ps) &$1.5535\pm0.0019$ &$1.5668\pm0.0012$ & 1.540 \\
$\Delta m$~(ps$^{-1}$) &$0.5085\pm0.0009$ &$0.4826\pm0.0006$ & 0.489\\
\Are(\rm comb)  &$0.0009\pm0.0004$ &$0.0085\pm0.0002$ &$0.0095\pm0.0002$  \\
\Arm(\rm comb)  &$0.0024\pm0.0005$ &$0.0103\pm0.0002$ &$0.0102\pm0.0002$  \\
\tBz(\rm comb)~(ps) &$1.3132\pm0.0017$ &$1.2898\pm0.0012$ &  \\
$\Delta m$(\rm comb)~(ps$^{-1}$) &$0.4412\pm0.0008$ &$0.4000\pm0.0005$ & \\
\hline
\end{tabular}
\end{center}
\end{table*}
The value of $\Delta m$ is consistent with the world
average, while
the value of \tBz\ is slightly larger than expected, an effect also
observed in the simulation. 
By fixing its value to the world average, the $\dCP$ result decreases by
$0.18\times 10^{-3}$. This effect is taken into account in the 
systematic uncertainty computation.
Figures \ref{f:Dz} and \ref{f:costhe} show the fit projections for \DT\
and cos$\theta_{\ell K}$, respectively.  

A sizeable charge asymmetry is observed at the reconstruction stage, for both
$e$ and $\mu$ reconstruction and at the \kaon\ tagging stage, somewhat
smaller than that observed in the simulation. 
As the size of \Ar\ is the same for the $e$ and the $\mu$ samples, it
is reasonable to suppose that the 
main source of charge asymmetry at the \Bz\ reconstruction stage is due to the
\psoft .  

Recently the \babar\  collaboration published a measurement of the asymmetry $\All$ between same-sign inclusive dilepton samples $\ell^+ \ell^+$ and $\ell^- \ell^-$ using the complete recorded data set~\cite{BaBar_dilep2}.
The systematic errors of the two analyses are essentially uncorrelated.
The correlation of the statistical errors is estimated to be on a level below 10 percent.

\section{Conclusions}
\label{sec:conc}
We present a new precise measurement of the parameter governing
\CP violation in {\Bz - \Bzb} oscillations. With a technique based on
partial \BtoDs\ reconstruction and \kaon\ tagging we find 
\begin{linenomath}
$$\dCP = 1- |q/p|=(0.29\pm0.84^{+1.88 }_{-1.61})\times 10^{-3},$$
\end{linenomath}
where the first uncertainty is statistical and the second is
systematic.
The corresponding asymmetry,
\begin{linenomath}
$$ \All \simeq
2(1 -  | \frac{q}{p}|) =(0.06\pm0.17^{+0.38}_{-0.32})\%,$$ is consistent with and
\end{linenomath}
competitive with the results from dilepton measurements
at the {\em B} factories{, LHCb~\cite{LHCB_Bd} and $D\emptyset$~\cite{D0_mumu}}. 
We observe no deviation from the SM expectation~\cite{SM}.

\begin{center} 
\begin{figure}[htbp]
\includegraphics[width=9.cm]{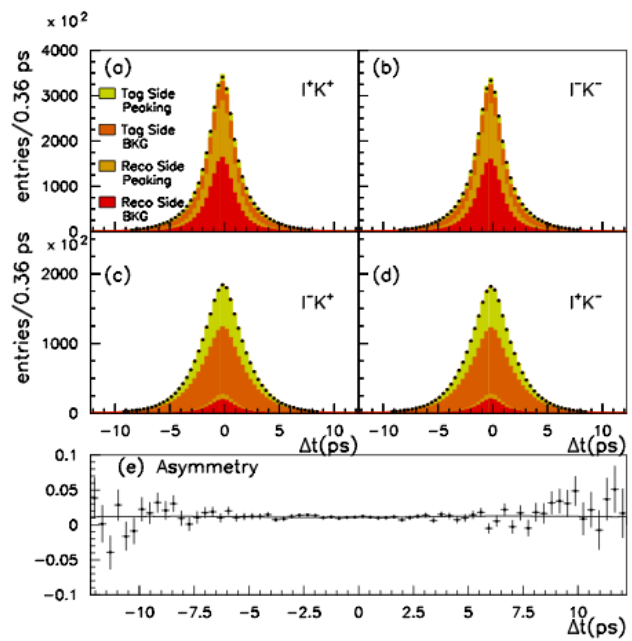}  
\caption{(color online): distribution of \DT\ for the
continuum-subtracted data (points with error bars) and fitted contributions
from peaking \Kt, background \Kt, peaking \Kr and background \Kr, for (a) $\ell^+ K^+$ events, (b)
$\ell^- K^-$ events, (c) $\ell^- K^+$ events, (d) $\ell^+ K^-$ events,
(e) raw asymmetry between $\ell^+ K^+$ and $\ell^- K^-$ events.}
\label{f:Dz}
\end{figure}
\end{center}

\begin{center} 
\begin{figure}[htbp]
\includegraphics[width=9.cm]{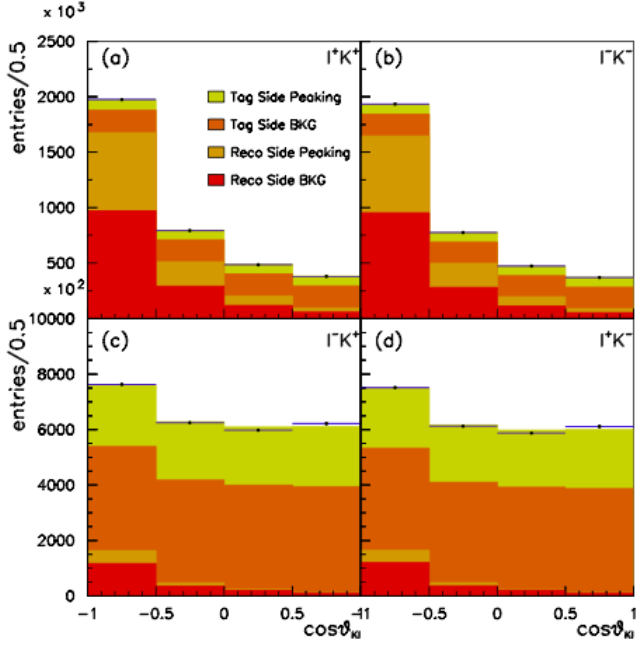}  
\caption{(color online): distribution of $\cos\theta_{\ell K}$ for the
continuum-subtracted data (points with error bars) and fitted contributions
from peaking \Kt, background \Kt, peaking \Kr and background \Kr, for (a) $\ell^+ K^+$ events, (b)
$\ell^- K^-$ events, (c) $\ell^- K^+$ events, (d) $\ell^+ K^-$ events.} 
\label{f:costhe}
\end{figure}
\end{center}

\section{Acknowledgments}
We are grateful for the 
extraordinary contributions of our \pep2\ colleagues in
achieving the excellent luminosity and machine conditions
that have made this work possible.
The success of this project also relies critically on the 
expertise and dedication of the computing organizations that 
support \babar.
The collaborating institutions wish to thank 
SLAC for its support and the kind hospitality extended to them. 
This work is supported by the
US Department of Energy
and National Science Foundation, the
Natural Sciences and Engineering Research Council (Canada),
the Commissariat \`a l'Energie Atomique and
Institut National de Physique Nucl\'eaire et de Physique des Particules
(France), the
Bundesministerium f\"ur Bildung und Forschung and
Deutsche Forschungsgemeinschaft
(Germany), the
Istituto Nazionale di Fisica Nucleare (Italy),
the Foundation for Fundamental Research on Matter (The Netherlands),
the Research Council of Norway, the
Ministry of Education and Science of the Russian Federation, 
Ministerio de Economia y Competitividad (Spain), and the
Science and Technology Facilities Council (United Kingdom).
Individuals have received support from 
the Marie-Curie IEF program (European Union) and the A. P. Sloan Foundation (USA).

\end{document}